\documentstyle[prl,aps,floats,psfig]{revtex}
\begin{document}
\twocolumn[
\hsize\textwidth\columnwidth\hsize\csname@twocolumnfalse\endcsname
\draft
\title{$c$-Axis Penetration Depth in the Cuprates:\\
Additional Evidence for Incoherent Hopping}
\author{R. J. Radtke and V. N. Kostur\cite{na}}
\address{Center for Superconductivity Research, Department of Physics,\\
University of Maryland, College Park, Maryland, 20742-4111}
\date{{\bf E-print cond-mat/9509nnn}, received \today}
\maketitle
\begin{abstract}
Measurements of the $c$-axis penetration depth $\lambda_c$ in the
cuprates reveal a low-temperature $T$ dependence which is inconsistent
with simple models of coupling between the CuO$_2$ layers.
In this paper, we examine whether a model based on {\it incoherent}
hopping between the layers can account for this low-$T$ behavior.
We compute $\lambda_c$ directly from linear response theory and
compare our results with recent experimental measurements
on $\rm YBa_2u_3O_{7-\delta}$ as a function of temperature and doping.
We find that the data can be reproduced within this model
providing the inter-layer  scattering is anisotropic and
the pairing is $d$-wave.
In addition, our calculations demonstrate that $1/\lambda_c^2$
is proportional to the $c$-axis critical current, which seems to be a
generic feature in weakly coupled layered superconductors.
\end{abstract}
\pacs{ }
]
{\it Introduction.}
Studies of the $c$-axis properties of the cuprate superconductors
have revealed many unusual features in both the normal and
superconducting states.\cite{Cooper}
In the normal state, the $c$-axis resistivity $\rho_c (T)$
can show either a
metallic or a semiconducting temperature dependence as a function of
the doping, while the in-plane resistivity is always metallic.\cite{Cooper}
In the superconducting state, strong evidence exists in a variety of
cuprates that the CuO$_2$ planes are coupled by the Josephson effect,
in contrast to the behavior of conventional
superconductors.\cite{KM,Shibauchi,Basov}
Additionally, $c$-axis penetration depth measurements
on $YBa_2Cu_3O_{7-\delta}$ (YBCO) crystals have
become available\cite{Anlage,Bonn} which agree with neither the
conventional theory nor with each other.

Recent work has been directed toward accounting for these anomalous
features within a theoretical framework based on incoherent quasiparticle
hopping between the CuO$_2$ planes.\cite{Cooper}
In this picture, the coupling between nearest-neighbor CuO$_2$
planes is so weak that quasiparticle transport occurs through
uncorrelated tunneling (or hopping) of the quasiparticles.
The origin of this incoherence is unclear, but it may be related to
strong intra-layer scattering,\cite{KumarJ} a non-Fermi-liquid ground
state within the layers,\cite{AZ}
dynamic inter-layer scattering,\cite{KumarL} or
strong electromagnetic fluctuations.\cite{Leggett}
Based on these ideas, several phenomenological models have been
developed to study the implications of this incoherence\cite{Graf,Rojo}
and have had some success in describing both $\rho_c (T)$\cite{Rojo,RLL}
and the $c$-axis critical current $j_c$.\cite{RLL,RL}

In this paper, we extend the calculations of
Refs.~\onlinecite{Rojo,RLL,RL} by deriving expressions for the
electromagnetic penetration depth along the
$c$-axis directly from linear response theory and compare our
results with recent experiments.\cite{Bonn}
Following Ref.~\onlinecite{Rojo}, we postulate the existence of three
mechanisms for quasiparticle transport along the $c$-direction:
a direct hopping induced by overlap of the quasiparticle wave functions,
impurity-assisted hopping due to disorder in the intercalating
layers separating the CuO$_2$ planes, and boson-assisted
hopping due to, for example, certain optical phonons in the cuprates.
We find that the inverse $c$-axis penetration depth $1 / \lambda_c$ is
the sum in quadrature of $1 / \lambda_c$ due to each mechanism
in isolation, and we derive expressions for these terms.
Our calculations demonstrate that $1/\lambda_c^2 \propto j_c$ as
hypothesized earlier\cite{RLL,RL} based on calculations within the
Lawrence-Doniach model,\cite{Clem} indicating that
this Josephson-like relation between $j_c$ and $\lambda_c$ is
a general property of weakly coupled layered superconductors.
Finally, we examine the temperature dependence of the contributions
of each interlayer hopping process to the penetration depth for
both $s$- and $d$-wave superconductors.
Comparing with the data of Bonn {\it et al.},\cite{Bonn} we find
that the doping and temperature dependence of the experimental
$\lambda_c^2 (0) / \lambda_c^2 (T)$ is qualitatively consistent
with $d$-wave pairing and anisotropic interlayer scattering.

{\it Penetration depth in the incoherent hopping model.}
To begin, consider electrons on a three-dimensional tetragonal
lattice governed by the Hamiltonian $H_{\rm el} = \sum_m H_m + H_{\bot}$.
The lattice is defined by the primitive vectors
$a {\bf \hat{x}}$, $a {\bf \hat{y}}$,
and $d {\bf \hat{z}}$ with $d \gg a$, so that it resembles a stack of layers,
which we index by $m$.
$H_m$ is the Hamiltonian for the motion in the $m$th layer and
$H_{\bot}$ describes the interlayer coupling.
For the purposes of this paper, it is not necessary to specify
$H_m$ completely; we need only assume that it
becomes a BCS-type superconductor below a
critical temperature $T_c$.
The interlayer coupling Hamiltonian may then be written in
terms of the annihilation and creation operators for the
quasiparticles obtained from $H_m$ at site $i\equiv (i_x,i_y)$ in layer $m$
with spin projection $\sigma$,
$c_{im\sigma}$ and $c_{im\sigma}^{\dag}$:\cite{Rojo,RL}
\begin{equation}
H_{\bot} = \sum_{im\sigma} t_{im}
  \left[ c_{i,m+1,\sigma}^{\dag} c_{im\sigma}^{~} + {\rm h.c.} \right]
\>,
\label{eq:Hbot}
\end{equation}
where
\begin{equation}
t_{im} = t_{\bot} + V_{im} + \sum_j g_{i-j,m}\phi_{jm}\>.
\label{eq:tim}
\end{equation}
The terms in this Hamiltonian represent interlayer hopping due
to overlap of the quasiparticle wave functions (parameterized by $t_{\bot}$),
impurity scattering (modeled by the random variable $V_{im}$), and
bosonic scattering (written in terms of the bosonic field operator
$\phi_{im}$ and its coupling strength to the quasiparticles $g_{im}$).
\cite{ref1}

As discussed in the Introduction, a significant number of
experiments suggest that the interlayer coupling in the cuprates is in
general weak and in particular incoherent.
A fully microscopic theory of this incoherence is beyond the
scope of this article, but its effects can be simulated by
performing calculations to second order in $H_{\bot}$.
This approach is similar in spirit to that of Ioffe {\it et al.}\cite{Ioffe}
and has been used successfully to study the normal-state
resistivity\cite{Rojo,RLL} and the $c$-axis critical current.\cite{RLL,RL}
We take this approach in what follows and refer the interested
reader to Ref.~\onlinecite{RL} for a more detailed description of
the perturbation theory and the assumptions underlying it.

To compute $\lambda_c$ within this incoherent hopping model,
we examine the current response to a weak vector potential
$A_{im} (t) {\bf \hat{z}}$ in linear respone theory.
As described by Peierls,\cite{Peierls}
the presence of the vector potential induces a phase
$e^{i (ed / \hbar c) A_{im}}$ into Eq.~(\ref{eq:Hbot}), where
$d$ is the inter-layer spacing, $e$ is the electronic charge,
and $c$ is the speed of light.
For weak fields, we may expand the exponential to second order
in $A_{im}$ and thereby obtain an expression for the total
charge current density:\cite{SWZ}
$j_{im}^{z,\rm Tot} = j_{im}^{z,\rm P} + j_{im}^{z,\rm D}$,
where the paramagnetic current
\begin{equation}
j_{im}^{z,\rm P} = - \frac{ie}{\hbar a^2} \sum_{\sigma}
  t_{im} \left[ c_{i,m+1,\sigma}^{\dag} c_{im\sigma}^{ } -
{\rm h.c} \right]\>,
\label{eq:jp}
\end{equation}
and the diamagnetic current
\begin{equation}
j_{im}^{z,\rm D} = \frac{e^2 d}{\hbar^2 a^2 c} \sum_{\sigma} t_{im} A_{im}
  \left[ c_{i,m+1,\sigma}^{\dag} c_{im\sigma}^{ } + {\rm h.c} \right]\>.
\label{eq:jd}
\end{equation}

In the usual way,\cite{Mahan} we expand the expectation value
of the total current to linear order in $A_{im}$, Fourier transform
the result in space and time, and use Maxwell's equations to
relate the electric field $E_{im} (t) {\bf \hat{z}}$ to the vector potential.
This procedure gives
$ j^{z,\rm Tot}_{\bf Q} (\omega)
  = \sum_{\bf Q'} \, \sigma_{\bf Q,Q'}^{zz} (\omega) \, E_{\bf Q'} (\omega)$,
where
\begin{equation}
\sigma_{\bf Q,Q'}^{zz} (\omega) =
  \frac{i}{\omega}
  \left[ \Pi_{\bf Q,Q'}^{zz} (\omega) + D_{\bf Q,Q'}^{zz} \right]
\label{eq:sigmaQQ}
\end{equation}
is the conducivity with
\begin{equation}
\Pi_{\bf Q,Q'}^{zz} (\omega) =
  - \frac{i}{\hbar} \frac{a^2 d}{N} \int_{-\infty}^{0} dt \, e^{i\omega t}
  \left< \left[ j^{z,\rm P}_{\bf Q} (t), j^{z,\rm P}_{\bf -Q'} (0)
\right] \right>
\label{eq:pi}
\end{equation}
proportional to the Fourier transform of a retarded current-current
correlation function and
\begin{eqnarray}
D_{\bf Q,Q'}^{zz} & = &
  - \frac{e^2 d}{\hbar^2 a^2} \frac{1}{N} \sum_{im\sigma}
  e^{-i({\bf Q} - {\bf Q'}) \cdot {\bf R}_{im}}\nonumber \\
& \times &  \left<  t_{im}
  \left[ c_{i,m+1,\sigma}^{\dag} c_{im\sigma}^{~} + {\rm h.c.}
 \right] \right>
\label{eq:d}
\end{eqnarray}
the diamagnetic term.
In these expressions, $N$ is the total number of sites in the lattice,
$ j^{z,\rm P}_{\bf Q} = \sum_{im} e^{-i{\bf Q \cdot R}_{im}} j^{z,\rm P}_{im}$,
${\bf Q} = ({\bf q}, q_z)$,
and ${\bf R}_{im} = ({\bf r}_{i},md)$ is the position vector of the lattice
site indexed by $im$.
(Throughout this paper, we use capital letters to denote 3D vectors
and small letters to denote 2D vectors).
Note that the conductivity is not diagonal in wave vector due
to the impurity term; after impurity averaging, translational
invariance will be restored and we will find
$\sigma_{\bf Q,Q'}^{zz} (\omega) \propto \delta_{\bf Q,Q'}$.

To proceed further we must account for the incoherent
nature of charge transport along the $c$-axis by expanding
both $\Pi_{\bf Q,Q'}^{zz}$ and $D_{\bf Q,Q'}^{zz}$ to second
order in the interlayer hopping amplitude $t_{im}$.
Thus, the paramagnetic current-current correlation
function, evaluated using the standard diagrammatic techniques
in Matsubara space, is represented by a {\it bare} particle-hole
bubble (vertex corrections being of higher order in $t_{im}$)
with purely {\it intra}-layer Green's functions renormalized only
by the {\it intra}-layer self energy.\cite{RL}
Although the diamagnetic term does not play any substantial role in
the optical conductivity because it is independent of frequency,
it is important for the penetration depth calculation and must be
treated carefully.
In fact, evaluating this term perturbatively yields a form nearly
identical to $\Pi_{\bf Q,Q'}^{zz}$ except for factors at the vertices,
as we shall see below.
Working in Matsubara space with Nambu Green's functions
for the intra-layer propagators\cite{AM} and assuming that
the layers $m$ are identical, we obtain
\begin{eqnarray}
\Pi_{\bf Q,Q'}^{zz} (i\nu_n) & = &
  \frac{2 e^2 d}{\hbar^2 a^2} \, \frac{\delta_{q_z,q_z'}}{N_{\|}^3}
\sum_{\bf k,k'}
  \int_{0}^{\beta} d\tau \, e^{i\nu_n \tau} \, \nonumber \\
  & \times & \left< T_{\tau} t_{\bf k + q - k'} (\tau)
 t_{\bf -k - q' + k'} \right> \,
\nonumber \\
  & \times & {\rm Tr} \left[ \hat{G}_{\bf k'} (\tau)
\hat{G}_{\bf k} (-\tau) \right]
\label{eq:Pin}
\end{eqnarray}
and
\begin{eqnarray}
D_{\bf Q,Q'}^{zz} & = &
  -\frac{2 e^2 d}{\hbar^2 a^2} \, \frac{\delta_{q_z,q_z'}}{N_{\|}^3}
\sum_{\bf k,k'}
  \int_{0}^{\beta} d\tau \, \nonumber \\
  & \times & \left< T_{\tau} t_{\bf k - k'} (\tau)
t_{\bf -k + q - q' + k'} \right> \,
\nonumber \\
  & \times & {\rm Tr} \left[ \hat{G}_{\bf k'} (\tau) \hat{\tau}_3
 \hat{G}_{\bf k} (-\tau) \hat{\tau}_3 \right] ,
\label{eq:Dn}
\end{eqnarray}
where $N_{\|}$ is the number of lattice site in a single layer,
$\hat{\tau}_3$ is the third Pauli matrix, and
$t_{\bf q} = \sum_{i} e^{-i{\bf q \cdot r}_{i}} t_{im}$ is
assumed to be independent of $m$.
($k_B = 1$ throughout this paper, and the rest of the notation
is standard.)

The correlation function $\left< T_{\tau} t_{\bf q} (\tau) t_{\bf q'} \right>$
represents a combination of an impurity average over $V_{im}$
and a thermodynamic average over the bosonic field $\phi_{im}$.
Since we go to second order in the hopping amplitudes and
take the mean of $V_{im}$ to be zero, this correlation function
decomposes into a sum of three components:  a direct hopping
term, an impurity-assisted hopping term,
and a boson-assisted hopping term.
The conductivity is therefore a sum of three terms, as obtained
previously.\cite{RL}
Since the penetration depth is given by
\begin{eqnarray}
\frac{c^2}{4\pi\lambda_c^2} & = & \lim_{{\bf Q} \rightarrow 0} \,
 \lim_{\omega \rightarrow 0} \,
  \left[ \omega \, {\rm Im} \, \sigma_{\bf Q,Q}^{zz} (\omega) \right]
\nonumber \\
& = & \lim_{\omega \rightarrow 0} \,
\lim_{{\bf Q} \rightarrow 0} \,
\left[ \omega \, {\rm Im} \, \sigma_{\bf Q,Q}^{zz} (\omega) \right]
\label{eq:lambdac}
\end{eqnarray}
after impurity averaging, this decomposition allows us to write
\begin{equation}
\left( \frac{1}{\lambda_c^2} \right)_{\rm Total} =
  \left( \frac{1}{\lambda_c^2} \right)_{\rm direct} +
  \left( \frac{1}{\lambda_c^2} \right)_{\rm imp} +
  \left( \frac{1}{\lambda_c^2} \right)_{\rm inel}
\end{equation}
and consider each term separately.

{\it Direct contribution.}
For the direct term, $t_{im} \rightarrow t_{\bot}$ [cf. Eq.~(\ref{eq:tim})],
from which
\begin{equation}
\left. \left<T_{\tau} t_{\bf q} (\tau) t_{\bf q'} \right>
\right|_{\rm direct} =
  N_{\|}^2 \delta_{\bf q,0} \delta_{\bf q',0} \, t_{\bot}^2 .
\end{equation}
Inserting this relation into Eqs.~(\ref{eq:Pin})-(\ref{eq:Dn}) gives
\begin{eqnarray}
\left. \Pi_{\bf Q,Q'}^{zz} (i\nu_n) \right|_{\rm direct} & = &
  \delta_{\bf Q,Q'} \, \frac{2 e^2 d}{\hbar^2 a^2} \, t_{\bot}^2 \,
  \frac{T}{N_{\|}}
\nonumber \\
& \times & \sum_{{\bf k}l}
  {\rm Tr} \left[ \hat{G}_{\bf k+q} (i\omega_{l+n}) \hat{G}_{\bf k}
(i\omega_l) \right]
\label{eq:Pi_direct}
\end{eqnarray}
and
\begin{eqnarray}
\left. D_{\bf Q,Q'}^{zz} \right|_{\rm direct} & = &
  - \delta_{\bf Q,Q'} \, \frac{2 e^2 d}{\hbar^2 a^2} \, t_{\bot}^2 \,
  \frac{T}{N_{\|}}
\nonumber \\
& \times & \sum_{{\bf k}l}
  {\rm Tr} \left[ \hat{G}_{\bf k} (i\omega_{l}) \hat{\tau}_3
  \hat{G}_{\bf k} (i\omega_l) \hat{\tau}_3 \right].
\label{eq:D_direct}
\end{eqnarray}
The resulting conductivity is, as expected, diagonal in wave vector.
Inserting Eqs.~(\ref{eq:Pi_direct})-(\ref{eq:D_direct}) into
Eq.~(\ref{eq:lambdac}), we have
\begin{equation}
\left( \frac{c^2}{4\pi\lambda_c^2} \right)_{\rm direct} =
  \frac{8 e^2 d}{\hbar^2 a^2} \, t_{\bot}^2 \, \frac{T}{N_{\|}}
\sum_{{\bf k}l}
  F_{\bf k} (i\omega_l) F_{\bf k} (i\omega_l),
\label{eq:lambda_direct}
\end{equation}
where $F_{\bf k} (i\omega_l)$ is the Gor'kov propagator.\cite{AM}

Several features of this result are noteworthy.
First and foremost, Eq.~(\ref{eq:lambda_direct}) has the same form as
the expression for the $c$-axis critical current; specifically,
$(1 / \lambda_c^2)_{\rm direct} \propto j_c^{\rm direct}$.\cite{RL}
As we shall see, this conclusion holds for the other two processes as
well.
This behavior arises from the cancellation of the normal
($\hat{\tau}_0$ and $\hat{\tau}_3$)
components of the quasiparticle propagators between the paramagnetic
and diamagnetic terms.
This cancellation is a special property of the $c$-axis
charge transport in weakly coupled layered superconductors
and reproduces the results obtained by Bulaevskii and Clem based on
the Lawrence-Doniach model.\cite{Clem}

Second, if we take the BCS form for the Gor'kov propagators
$F_{\bf k} (i\omega_l) = \Delta_{\bf k} / ((i\omega_l)^2 - E_{\bf k}^2)$,
in the limit where the superconducting energy scales are smaller than
the intra-layer electronic energy scales, we obtain
\begin{equation}
\left( \frac{1}{\lambda_c^2} \right)_{\rm direct} =
  \frac{16 \pi^2 e^2 d}{\hbar^2 c^2} \, N(0) t_{\bot}^2 \, T \sum_l
  \left< \frac{\Delta_{\bf k}^2}{\left[ \omega_l^2 + \Delta_{\bf k}^2
\right]^{3/2}}
  \right>_{\bf k},
\label{eq:lambda_direct_fs}
\end{equation}
where $N(0)$ is two-dimensional density of states at the Fermi surface
(for a parabolic spectrum $\epsilon_{\bf k}= \hbar^2 k^2/2m - E_F$,
$N(0) = m/2\pi\hbar^2$),
$\Delta_{\bf k}$ is the superconducting gap function,
$E_{\bf k} = \sqrt{\epsilon_{\bf k}^2 + \Delta_{\bf k}^2}$,
and
the angle brackets denote a normalized Fermi surface average over
the indicated variable.
Comparing with the standard BCS result,\cite{AGD} we see that this
term has the same temperature dependence but a different
magnitude.\cite{RL}
This behavior arises from a combination of the dimensionality of the
layers and the conservation of the transmitted wave vector.\cite{RL}

{\it Impurity-assisted contribution.}
The impurity-assisted term is computed in an analogous way.
In this case, $t_{im} \rightarrow V_{im}$ [cf. Eq.~(\ref{eq:tim})] and the
impurity average yields $\overline{V_{im} V_{jm}} = \overline{V^2}_{i-j}$.
The correlation function in Eqs.~(\ref{eq:Pin})-(\ref{eq:Dn}) is then
\begin{equation}
\left. \left< T_{\tau} t_{\bf q} (\tau) t_{\bf q'} \right>
\right|_{\rm imp} =
  N_{\|} \delta_{\bf q,-q'} \, \overline{V^2}_{\bf q},
\end{equation}
from which we obtain
\begin{eqnarray}
\left. \Pi_{\bf Q,Q'}^{zz} (i\nu_n) \right|_{\rm imp} & = &
  \delta_{\bf Q,Q'} \, \frac{2 e^2 d}{\hbar^2 a^2} \,
  \frac{T}{N_{\|}^2} \sum_{{\bf kk'}l} \overline{V^2}_{\bf k - k'} \,
\nonumber \\
  & \times & {\rm Tr} \left[ \hat{G}_{\bf k' + q}
(i\omega_{l+n}) \hat{G}_{\bf k} (i\omega_l) \right]
\label{eq:Pi_imp}
\end{eqnarray}
and
\begin{eqnarray}
\left. D_{\bf Q,Q'}^{zz} \right|_{\rm imp} & = &
  - \delta_{\bf Q,Q'} \, \frac{2 e^2 d}{\hbar^2 a^2} \,
  \frac{T}{N_{\|}^2} \sum_{{\bf kk'}l} \overline{V^2}_{\bf k - k'} \,
\nonumber \\
  & \times & {\rm Tr} \left[ \hat{G}_{\bf k'} (i\omega_{l}) \hat{\tau}_3
  \hat{G}_{\bf k} (i\omega_l) \hat{\tau}_3 \right].
\label{eq:D_imp}
\end{eqnarray}
The penetration depth follows from these relations as
\begin{equation}
\left( \frac{c^2}{4\pi\lambda_c^2} \right)_{\rm imp} =
  \frac{8 e^2 d}{\hbar^2 a^2} \, \frac{T}{N_{\|}^2} \sum_{{\bf kk'}l}
  \overline{V^2}_{\bf k - k'} \, F_{\bf k} (i\omega_l)
F_{\bf k'} (i\omega_l).
\label{eq:lambda_imp}
\end{equation}
Again, the penetration depth has the same form as the corresponding
critical current calculation:
$(1 / \lambda_c^2)_{\rm imp} \propto j_c^{\rm imp}$.\cite{RL}
In the BCS limit taken above, Eq.~(\ref{eq:lambda_imp}) may be written
\begin{eqnarray}
\left( \frac{1}{\lambda_c^2} \right)_{\rm imp} & = &
  \frac{32 \pi^3 e^2 v}{\hbar^2 c^2} \, N^2(0) \, T \sum_l
  \left< \left< \overline{V^2}_{\bf k - k'} \right. \right.
\nonumber \\
& \times &  \left. \left.
\frac{\Delta_{\bf k}}{\sqrt{\omega_l^2 + \Delta_{\bf k}^2}}
  \frac{\Delta_{\bf k'}}{\sqrt{\omega_l^2 + \Delta_{\bf k'}^2}}
  \right>_{\bf k} \right>_{\bf k'},
\label{eq:lambda_imp_fs}
\end{eqnarray}
where $v=a^2d$ is the unit cell volume.

The contribution of the impurity-assisted component to the
total penetration depth depends on both the pairing symmetry
and the anisotropy of the matrix element.
If the pairing is isotropic ($\Delta_{\bf k} = \Delta$), this
expression reduces to the Ambegaokar-Baratoff result for SIS
tunnel junctions.\cite{AB}
However, if the pairing is $d$-wave (or any pairing
state with $\left< \Delta_{\bf k} \right>_{\bf k} = 0$) and the
matrix element is isotropic ($\overline{V^2}_{\bf k - k'} =
\overline{V^2}$),
$\left( 1/\lambda_c^2 \right)_{\rm imp}$ vanishes by
symmetry.\cite{Graf,RL}
This result is a special consequence of an isotropic scattering matrix
element and does not hold when anisotropy, which should be
present in any real material, is included.

To see the effects of anisotropy on the impurity-assisted contribution
to the $c$-axis penetration depth, consider a scattering
matrix element of the form
\begin{equation}
\overline{V^2}_{\bf k - k'} =
  V^2 \, \frac{k_F\delta k}{({\bf k - k'})^2 + \delta k^2}.
\label{eq:imp_me}
\end{equation}
When $\delta k \rightarrow \infty$, $\overline{V^2}_{\bf k - k'}$
becomes isotropic, and we recover the results discussed above.
Physically, this limit corresponds to a total randomization of the
wave vector of the scattered quasiparticle and
is analogous to diffuse transmission through a tunnel junction.
In the opposite limit where $\delta k / k_F \rightarrow 0$, the
scattering matrix element preserves the direction of the
scattered quasiparticle and is analogous to specular transmission
through a tunnel junction.
In this limit, Eq.~(\ref{eq:lambda_imp_fs}) reduces to
a generalization of the Ambegaokar-Baratoff\cite{AB} result:
\begin{equation}
\left( \frac{\lambda_c^2 (0)}{\lambda_c^2 (T)} \right)_{\rm imp}
  \stackrel{\delta k \rightarrow 0}{\rightarrow}
  \left< \frac{\Delta_{\bf k}}{\Delta_{0} }
  \tanh \left( \frac{\Delta_{\bf k}}{2T} \right) \right>_{\bf k},
\label{eq:lambda_ab}
\end{equation}
where  $\Delta_{0}=\left<|\Delta_{\bf k} (T = 0) | \right>_{\bf k}$
is the average of the absolute value of the gap function over the
Fermi surface at zero temperature.
This form yields a finite contribution to the penetration depth
for {\it both} $s$- and $d$-wave pairing due to the
wave-vector-conserving nature of the scattering.

\begin{figure}
\centerline{\psfig{file=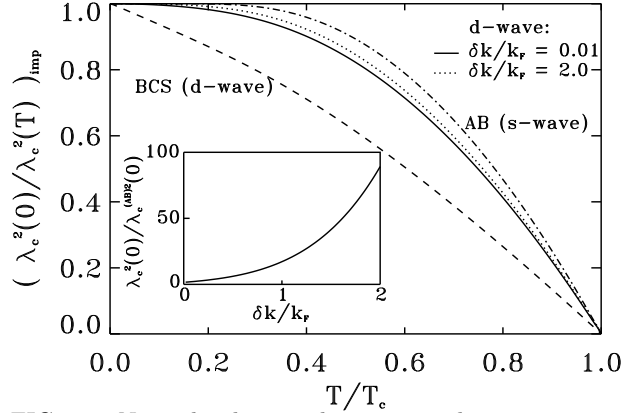,width=0.99\linewidth}}
\caption{Normalized contribution to the $c$-axis penetration depth
from impurity-assisted interlayer hopping
$\left( \lambda_c^2(0) / \lambda_c^2(T) \right)_{\rm imp}$
as a function of the reduced temperature $T / T_c$ in a $d$-wave
superconductor with slight ($\delta k / k_F$ = 0.01, solid line) and
significant ($\delta k / k_F$ = 2.0, dotted line)
non-conservation of wave vector [cf. Eq.~(\protect\ref{eq:imp_me})].
For comparison, the $d$-wave BCS (dashed line) and the $s$-wave
Ambegaokar-Baratoff\protect\cite{AB} results (dot-dashed line)
are also shown.
Inset:  Square of the $c$-axis penetration depth at zero
temperature $\lambda_c^2 (0)$ as a function of the spread in
transmitted wave vector $\delta k / k_F$
[cf. Eq.~(\protect\ref{eq:imp_me})] normalized to its value at
$\delta k / k_F$ = 0, $\lambda_c^{(AB)2}(0)$.}
\label{fig:imp}
\end{figure}
For intermediate $\delta k / k_F$, we can evaluate
$\left( 1 / \lambda_c^2 \right)_{\rm imp}$ from
Eqs.~(\ref{eq:lambda_imp_fs})-(\ref{eq:imp_me}) for a $d$-wave
superconductor with the results shown in Fig.~\ref{fig:imp}.
At zero temperature, we see from the inset to this figure that
the magnitude of the penetration depth depends strongly on
$\delta k / k_F$, in keeping with the fact that
$(\lambda_c^{-2} (0))_{\rm imp}$
vanishes in the $\delta k \rightarrow \infty$ limit.
As seen from the main figure, however, the temperature dependence
of the penetration depth ratio is {\it not} strongly affected by the amount
of scattering anisotropy:
the penetration depth ratios for $\delta k / k_F$ = 0.01 and 2.0 lie
very nearly on top of each other.
Moreover, we see that the temperature dependence of the anisotropic
impurity-assisted scattering is distinguishable from both the
conventional BCS result and the $s$-wave
Ambegaokar-Baratoff result.

Our calculations therefore suggest two experimental signatures
of anisotropic impurity-assisted interlayer scattering:
(1) the magnitude of the $c$-axis penetration depth should depend
strongly on the type and quantity of disorder, but the temperature
dependence should be largely insensitive to these factors, and
(2) the temperature dependence of the penetration depth ratio should
be between the BCS and Ambegaokar-Baratoff predictions.

{\it Boson-assisted contribution.}
The boson-assisted hopping term is more involved due to the
inelastic scattering (i.e., the retardation effects)
induced by the boson.
The hopping amplitude $t_{im} \rightarrow \sum_j g_{i-j,m} \phi_{jm}$
[cf. Eq.~(\ref{eq:tim})], and so the thermodynamic average returns a factor
of the boson propagator
$D_{{\bf q}m} (\tau) =
  - \sum_{ij} e^{-i{\bf q \cdot} ({\bf r}_{i}-{\bf r}_{j})} \,
  \left< T_{\tau} \left[ \phi_{im} (\tau) \phi_{jm} (0) \right] \right>$,
allowing the correlation function in Eqs.~(\ref{eq:Pin})-(\ref{eq:Dn})
to be written
\begin{equation}
\left. \left< T_{\tau} t_{\bf q} (\tau) t_{\bf q'} \right>
\right|_{\rm inel} =
  - N_{\|} \delta_{\bf q,-q'} \, \left| g_{{\bf q}m} \right|^2
D_{{\bf q}m} (\tau).
\end{equation}
If the layers are identical, we may drop the index $m$ and compute
the conductivity and the penetration depth as in the preceding two
cases, giving
\begin{eqnarray}
\left. \Pi_{\bf Q,Q'}^{zz} (i\nu_n) \right|_{\rm inel} & = &
  - \delta_{\bf Q,Q'} \, \frac{2 e^2 d}{\hbar^2 a^2} \,
  \frac{T^2}{N_{\|}^2}
 \nonumber \\
 & \times & \sum_{{\bf kk'}ll'}
  \left| g_{\bf k - k'} \right|^2  D_{\bf k - k'} (i\nu_{l - l'}) \,
\nonumber \\
& \times &  {\rm Tr} \left[ \hat{G}_{\bf k' + q} (i\omega_{l+n})
\hat{G}_{\bf k} (i\omega_l) \right],
\label{eq:Pi_inel}
\end{eqnarray}
\begin{eqnarray}
\left. D_{\bf Q,Q'}^{zz} \right|_{\rm inel} & = &
  \delta_{\bf Q,Q'} \, \frac{2 e^2 d}{\hbar^2 a^2} \,
  \frac{T^2}{N_{\|}^2}
\nonumber \\
 & \times & \sum_{{\bf kk'}ll'}
  \left| g_{\bf k - k'} \right|^2 D_{\bf k - k'} (i\nu_{l - l'}) \,
\nonumber \\
& \times &   {\rm Tr} \left[ \hat{G}_{\bf k'} (i\omega_{l'}) \hat{\tau}_3
  \hat{G}_{\bf k} (i\omega_l) \hat{\tau}_3 \right],
\label{eq:D_inel}
\end{eqnarray}
and
\begin{eqnarray}
\left( \frac{c^2}{4\pi\lambda_c^2} \right)_{\rm inel} & = &
  -\frac{8 e^2 d}{\hbar^2 a^2} \, \frac{T^2}{N_{\|}^2} \sum_{{\bf kk'}ll'}
  \left| g_{\bf k - k'} \right|^2 \, D_{\bf k - k'} (i\nu_{l - l'}) \,
\nonumber \\
  & \times & F_{\bf k} (i\omega_l) F_{\bf k'} (i\omega_{l'}).
\label{eq:lambda_inel}
\end{eqnarray}
As before, this expression is proportional to the corresponding one
for the critical current.\cite{RL}
In the BCS limit, we can introduce the spectral representation of the
boson propagator familiar from the theory of superconductivity,\cite{ref2}
\begin{equation}
B_{\bf k,k'} (\Omega) =
  - \frac{1}{\pi} \, \left| g_{\bf k - k'} \right|^2 \, {\rm Im}
D_{\bf k - k'} (\Omega),
\end{equation}
and write the result as
\begin{eqnarray}
\left( \frac{1}{\lambda_c^2} \right)_{\rm inel} &=&
  \frac{32 \pi^3 e^2 v}{\hbar^2 c^2} \, N^2(0) \,
  \left< \left< \int_0^{\infty} d\Omega \,B_{\bf {k,k'}} (\Omega)
  \right. \right. \nonumber \\
  &\times&   T^2 \sum_{ll'} \frac{2\Omega}{\nu_{l-l'}^2 + \Omega^2} \,
  \frac{\Delta_{\bf k}}{\sqrt{\omega_l^2 + \Delta_{\bf k}^2}}
\nonumber \\
  & \times & \left. \left.
\frac{\Delta_{\bf k'}}{\sqrt{\omega_{l'}^2 + \Delta_{\bf k'}^2}}
  \right>_{\bf k} \right>_{\bf k'}.
\label{eq:lambda_inel_fs}
\end{eqnarray}

For a simple Einstein phonon with a structureless coupling
constant
($B_{\bf k,k'} (\Omega) \propto \delta (\Omega - \Omega_0)$),
we find that $\left( 1 / \lambda_c^2 \right)_{\rm inel}$ resembles the
Ambegaokar-Baratoff form for $s$-wave pairing and
$\Omega_0 > \Delta_0$, but vanishes when the pairing is
$d$-wave.\cite{RL}
As with impurity-assisted scattering, however, anisotropy in
$B_{\bf k,k'} (\Omega)$ will in general be present
and will have a profound effect on
$\left( 1 / \lambda_c^2 \right)_{\rm inel}$
in $d$-wave superconductors.
Thus, we need to quantify the effects of anisotropy in this case.

The anisotropy in the boson-assisted scattering
may occur due to wave-vector dependence in either the
electron-boson coupling constant $g_{\bf q}$ or the dispersion
of the boson itself, so it is difficult to obtain a simple form for
$B_{\bf k,k'} (\Omega)$ similar to Eq.~(\ref{eq:imp_me}).
We therefore take an alternate approach and expand
the wave vector dependence of all quantities in
Eq.~(\ref{eq:lambda_inel_fs}) in terms of Fermi surface harmonics
$F_L ({\bf k})$.\cite{FSH}
We further simplify our results by taking an Einstein spectrum for
each component of $B$,
$B_{L,L'} (\Omega) =
  \lambda_{L,L'} \, (\Omega_{L,L'} / 2 N(0)) \,
  \delta (\Omega - \Omega_{L,L'})$.
This procedure gives
$(1/\lambda_c^2)_{\rm inel}  = \sum_{L,L'} (1/\lambda_c^2)_{L,L'}$
with
\begin{eqnarray}
\left( \frac{1}{\lambda_c^2} \right)_{L,L'} & = &
  \frac{32 \pi^3 e^2 v}{\hbar^2 c^2} \, N(0) T^2 \,\sum_{ll'}
  \frac{\lambda_{L,L'} \Omega_{L,L'}^2}{\nu_{l-l'}^2+\Omega_{L,L'}^2}
\nonumber \\
  & \times & A_L (i\omega_l) A_{L'} (i\omega_{l'})
\label{eq:lambda_inel_fsh}
\end{eqnarray}
and
$A_L (i\omega_l) = \left< F_L ({\bf k}) \Delta_{\bf k}
  / \sqrt{\omega_l^2 + \Delta_{\bf k}^2} \right>_{\bf k}$.
For concreteness, we take a cylindrical Fermi surface within each
layer, which gives $F_L ({\bf k}) = \sqrt{2} \cos (L \phi)$, where
${\bf k} = k_F (\cos \phi, \sin \phi)$.

We can make several general statements by looking at the form of
Eq.~(\ref{eq:lambda_inel_fsh}).
First, although the total $(1/\lambda_c^2)_{\rm inel}$
must be positive, the individual components $(1/\lambda_c^2)_{L,L'}$
may have either sign because of the difference in the signs
of $A_L$ and $A_{L'}$.
Second, for isotropic $s$-wave pairing
($\Delta_{\bf k} = \Delta$), only the $L = L' = 0$ component is non-zero,
and we recover the results of Ref.~\onlinecite{RL}
after analytic continuation.
For $d$-wave pairing, on the other hand, only components with
$(L,L') = (4n+2,4n'+2)$ ($n$ and $n'$ integers) are non-zero.

\begin{figure}
\centerline{\psfig{file=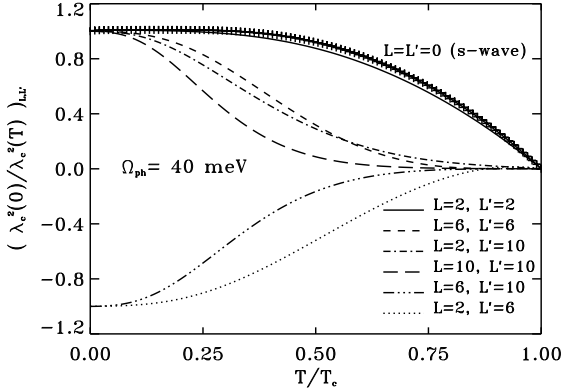,width=0.99\linewidth}}
\caption{Normalized matrix elements of the $c$-axis penetration
depth in the basis of Fermi surface harmonics
$\left( \lambda_c^{-2} (T) \right)_{L,L'}
  / \left| \left( \lambda_c^{-2} (0) \right)_{L,L'} \right|$ as
a function of the reduced temperature $T / T_c$ for a
$d$-wave superconductor and $(L,L')$ = (2,2) (solid line);
(6,6) (dashed line), (2,10) (dot-dashed line); (10,10) (long dashed line);
(6,10) (dot-dot-dot-dashed line); and (2,6) (dotted line).
Due to the symmetry of the the order parameter, only matrix
elements with $(L,L') = (4n + 2,4n' + 2)$, $n$ and $n'$
positive integers, are non-zero.
For comparison, the $s$-wave, $(L,L')$ = (0,0) result is plotted
as the thick solid line.
See text for details.}
\label{fig:inel}
\end{figure}

These features are exhibited in Fig.~\ref{fig:inel}, which shows
$(1/\lambda_c^2)_{L,L'}$ for both $s$- and $d$-wave pairing
computed numerically from Eq.~(\ref{eq:lambda_inel_fsh}).
In these calculations, we take $\Omega_{L,L'}=\Omega_{ph}$
= 40 meV to reflect
the expected importance of this phonon mode in $c$-axis
transport.\cite{Nyhus}
For isotropic $s$-wave pairing, only the $(L,L') = (0,0)$
component contributes, and we obtain a result which is similar to,
though slightly different from, the Ambegaokar-Baratoff result
as long as the boson involved has an energy scale larger than
the maximum of the gap function.\cite{RL}
For $d$-wave pairing, many $(L,L')$ components are non-zero
and some are negative, as suggested above.

Examining the $d$-wave curves more closely reveals several
interesting features.
At low temperatures, the non-zero components do not have the
linear-in-$T$ behavior typical of $d$-wave superconductors,
but seem to follow a higher power law.
This result is consistent with the fact that the nodes in a $d$-wave
gap function must give rise to a power law in the penetration
depth, but we see that the exact exponent for this power law
is determined by the mechanism responsible for the hopping and
the symmetry of its matrix elements.
At higher temperatures, the contribution from higher harmonics
($L$, $L' > 2$) vanishes.
This feature results from the orthogonality of the Fermi surface
harmonics in the following way.
As $T \rightarrow T_c$, the magnitude of $\Delta_{\bf k}$
decreases and
$A_L (i\omega_n) \sim
  \left< F_L ({\bf k}) \Delta_{\bf k} \right>_{\bf k} / |\omega_n|$.
Because the gap function is pure $d$-wave ($L$ = 2),
$\Delta_{\bf k} \propto F_2 ({\bf k})$ and so $A_L (i\omega_n)$,
and hence $\left( 1 / \lambda_c^2 \right)_{L,L'}$, goes to zero
when $L$, $L' > 2$ much faster than for the $L=L'$ = 2 component.
We have also computed the magnitudes of the
penetration depth components $\left( 1 / \lambda_c^2 \right)_{L,L'}$
and find that they depend strongly on $(L,L')$, becoming
considerably smaller with larger $L$ and $L'$.
Thus, one expects that the $(L,L') = (2,2)$ component will dominate
the boson-assisted contribution to the $c$-axis penetration depth
in a $d$-wave superconductor.

The conclusion one can draw from this analysis is that
boson-assisted hopping in a $d$-wave superconductor can contribute
to the $c$-axis penetration depth if the coupling is anisotropic or
if the boson has some dispersion.
Moreover, the resulting contribution to the penetration depth
is not necessarily linear in $T$ at low $T$, contrary to what one
expects and similar to what is observed experimentally.

\begin{figure}
\centerline{\psfig{file=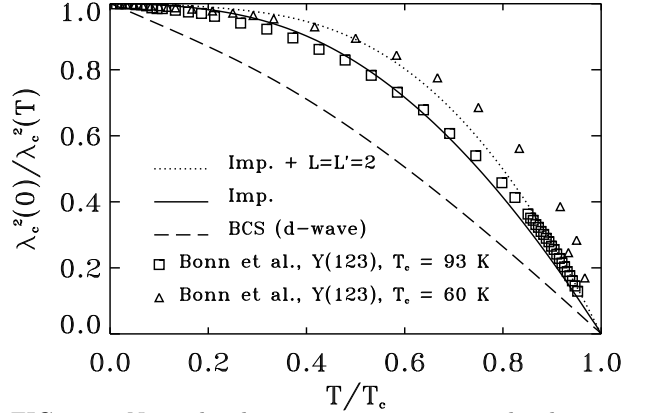,width=0.99\linewidth}}
\caption{Normalized $c$-axis penetration depth ratio
$\lambda_c^2(0) / \lambda_c^2(T)$ as a function of the
reduced temperature $T / T_c$ for a $d$-wave superconductor
with coherent (BCS-like) inter-layer coupling (dashed line),
incoherent coupling due to disorder-assisted hopping with
$\delta k / k_F$ = 0.01 (solid line; cf. Eq.~(\protect\ref{eq:imp_me})),
and incoherent hopping from a linear combination of 80 \%
disorder- ($\delta k / k_F$ = 0.01) with 20 \% boson-assisted hopping
(dotted line; cf. Eq.~(\protect\ref{eq:lambda_inel}) with $(L,L')$ = (2,2)).
Also shown are the data of Bonn {\it et al.}\protect\cite{Bonn}
for $T_c$ = 93 K (open squares) and $T_c$ = 60 K (open triangles)
$\rm YBa_2Cu_3O_{7-\delta}$.
Observe that effect of adding boson-assisted hopping is qualitatively
the same as de-oxygenation in the experimental data:
the penetration depth ratio is larger at larger temperatures.}
\label{fig:data}
\end{figure}

{\it Relation to experiment.}
In order to connect our results more firmly to experiment,
we compare some representative curves in our model to the
experimental $\lambda_c$ data of Bonn {\it et al.}\cite{Bonn}
on YBCO in Fig.~\ref{fig:data}.
The dashed line shows the BCS prediction for a $d$-wave
superconductor:  it is clearly inconsistent with these data.
The solid and dotted curves show the penetration depth ratio for
purely disorder-mediated hopping and a linear combination of disorder-
and boson-assisted hopping.
We see that the penetration depth ratio computed within the incoherent
hopping model is qualitatively consistent with these data:
despite having a $d$-wave order parameter, the low-temperature
behavior shows a much smaller slope, and the penetration depth ratio
falls much more slowly than BCS theory predicts.

We are also able to understand the systematic change in the
temperature dependence of the penetration depth ratio with
de-oxygenation.
Several experimental and structural features of YBCO,\cite{Cooper}
along with an analysis of its resistivity within our incoherent
hopping model,\cite{Rojo,RLL,RL} suggest that the fully oxygenated
compound is both ``cleaner'' and more three-dimensional than
the de-oxygenated samples.
In particular, we find that the boson-assisted hopping is more
pronounced in the de-oxygenated compound and is, in fact,
responsible for the upturn in the $c$-axis resistivity with
decreasing temperature.\cite{Rojo,RLL,RL}
We may therefore expect that the $c$-axis penetration depth ratio
will have a larger contribution due to the boson-assisted component
in the de-oxygenated samples while the disorder-assisted component
remains relatively constant.
The net result is that the penetration depth ratio will fall more
slowly with temperature in the de-oxygenated samples, and this
is what is observed experimentally.

{\it Summary.}
Viewing the CuO$_2$ layers in the cuprate superconductors
as incoherently coupled explains the features of
the anisotropic resistivity in these materials\cite{Rojo,RLL} and
makes definite predictions for the $c$-axis critical current
$j_c$.\cite{RL}
We have extended this theory by providing a microscopic
derivation of the $c$-axis penetration depth $\lambda_c$ and
find that $j_c \propto 1 / \lambda_c^2$, which seems to be a
generic property of superconductors consisting
of weakly coupled layers.\cite{Clem}
We have also examined the contributions to $\lambda_c$ arising
from direct, impurity-assisted, and boson-assisted hopping.
We find that anisotropy in the scattering matrix elements or
boson dispersion are required in order to obtain a finite penetration
depth in $d$-wave superconductors in the absence of direct
hopping, although no such restriction arises for $s$-wave pairing.
By computing the temperature dependence of the penetration depth
ratio $\lambda_c^2 (0) / \lambda_c^2 (T)$ for $d$-wave pairing
and anisotropic scattering, we find that the experimental
$c$-axis penetration depth ratio as a function of both temperature
and doping is qualitatively consistent with this model
for at least some measurements of $\lambda_c$ in YBCO.

\acknowledgements

The authors would like to thank D. A. Bonn and S. Kamal for making
their penetration depth data available to us prior to publication
and K. Levin for valuable discussions.
This work was supported by the NSF (RJR) and NASA (VK).


\begin{references}
\bibitem{ref1} For simplicity, we ignore interlayer hopping processes
between different intra-layer sites $i$ represented by terms like
$c_{i,m+1,\sigma}^{\dag} c_{i+\delta,m,\sigma}^{~}$
($\delta$ is the position vector to the nearest-neighbor sites).
One can extend the analysis presented here to include these terms,
but the results are unchanged as long as the interlayer spacing is
much larger than the intra-layer nearest-neighbor spacing.
\end{references}
\end{document}